\documentclass[aps,prl,twocolumn,groupedaddress]{revtex4}  
\usepackage{graphicx}  
\usepackage{dcolumn}   
\usepackage{bm}        
\usepackage{amssymb}   

\newcommand{\eqref}[1]{{(\ref{#1})}}

\newcommand{\be}{\begin{equation}}
\newcommand{\ee}{\end{equation}}
\newcommand{\bea}{\begin{eqnarray}}
\newcommand{\eea}{\end{eqnarray}}

\newcommand{\ti}{\times}

\newcommand{\mc}{\mathcal}

\newcommand{\beqa}{\begin{eqnarray}}
\newcommand{\eeqa}{\end{eqnarray}}

\begin{document}


\title{Can black hole superradiance be induced by galactic plasmas?}
\author{Joseph P. Conlon}\email{joseph.conlon@physics.ox.ac.uk} \affiliation{Rudolf Peierls Centre for Theoretical Physics, 1 Keble Road, Oxford, OX1 3NP, UK}
\author{Carlos A. R. Herdeiro}\email{herdeiro@ua.pt}
 \affiliation{Departamento de F\'isica da Universidade de Aveiro and CIDMA,
 Campus de Santiago, 3810-183 Aveiro, Portugal}
\date{December 2017}

\begin{abstract}

Highly spinning Kerr black holes with masses $M = 1 - 100\  M_{\odot}$ are subject to an efficient superradiant instability in the presence of
bosons with masses $\mu \sim 10^{-10} - 10^{-12}\  {\rm eV}$. We observe that this matches
the effective plasma-induced photon mass in diffuse galactic or intracluster environments ($\omega_{\rm pl} \sim 10^{-10} - 10^{-12}\ {\rm eV}$).
This suggests that bare Kerr black holes within galactic or intracluster environments, possibly even
including the ones produced in recently observed gravitational wave events, are unstable to formation of a
photon cloud that may contain a significant fraction
 of the mass of the original black hole.
At maximal efficiency, the instability timescale for a massive vector is milliseconds, potentially leading to a transient
rate of energy extraction from a black hole
in principle as large as $\sim 10^{55} \ {\rm erg \, s}^{-1}$.
We discuss possible astrophysical effects this could give rise to, including a speculative connection to Fast Radio Bursts.
\end{abstract}

\maketitle

\section{Introduction}
The recent observations of gravitational waves by LIGO, interpreted as binary black hole (BH) mergers~\cite{Abbott:2016blz,Abbott:2016nmj,Abbott:2017vtc,Abbott:2017oio,Abbott:2017gyy}, provide evidence for the abundant existence of stellar mass BHs with $M \lesssim 70 M_{\odot}$, likely described by the Kerr metric~\cite{Kerr:1963ud}. It is well known that these rotating BHs can release their rotational energy via the remarkable phenomenon of~\textit{superradiance}~\cite{Brito:2015oca}. This mechanism amplifies modes of a bosonic particle which are sufficiently low frequency and co-rotate with the BH. Confining such modes within the vicinity of the Kerr BH, for instance due to the particle's mass $\mu$, triggers an instability~\cite{Press:1972zz}. The superradiant modes undergo repeated amplification, and the BH's rotational energy sources the growth of a boson cloud around the BH.

The endpoint of the process is unknown, despite recent progress with fully non-linear numerical simulations~\cite{East:2017ovw}. For a real boson field (unlike a complex one~\cite{Herdeiro:2014goa,Herdeiro:2016tmi,Herdeiro:2017phl}), there is no known stationary state of a BH with a boson cloud, and so the latter must decay. Depending on the effect of non-linearities, two possible scenarios are $(i)$ a smooth saturation of the growth, followed by a steady cloud decay via the emission of gravitational radiation~\cite{Arvanitaki:2009fg}; or $(ii)$ the occurance of an explosive phenomenon akin to a  \textit{bosenova} that generates higher frequency bosonic modes that are expelled from the cloud towards infinity~\cite{Arvanitaki:2010sy,Yoshino:2012kn}.

For test fields in the linear regime, the instability has been studied most extensively for the case of a massive scalar~\cite{Detweiler:1980uk,Zouros:1979iw,Dolan:2007mj,Rosa:2009ei,Hod:2012zza,Dolan:2012yt,Furuhashi:2004jk,Hod:2016iri}, but recently it has also been possible to obtain the growth rates for the case of a massive vector, using either approximate schemes~\cite{Pani:2012vp,Pani:2012bp,Baryakhtar:2017ngi} or time evolutions~\cite{Witek:2012tr,East:2017mrj}.

Superradiant instabilities triggered by a massive bosonic field are most efficient when the gravitational radius of the BH is of the order of the Compton wavelength of the particle. In natural units, this means the \textit{mass coupling} is order unity:
\be
M\mu\sim 1\ \  \Leftrightarrow  \ \
\left( \frac{M}{M_{\odot}} \right) \left( \frac{\mu}{10^{-10} {\rm \ eV}} \right) \sim 1 \ ,
\label{superradiancecondition}
\ee
where $M$ is the BH mass. Thus, for astrophysical stellar mass BHs ranging from $1-100$ $M_\odot$, an efficient instability is achieved for ultralight bosons
with masses $\mu \sim 10^{-10} - 10^{-12} \, {\rm eV}$. The case of supermassive BHs requires even lighter particles. Since
no fundamental particles in this mass range are known, many astrophysical studies of the superradiant instability invoke physics beyond the standard
model, $e.g.$ the axiverse scenario~\cite{Arvanitaki:2009fg}.

An alternative possibility, which does not rely on new physics, is that superradiance is caused by an \textit{effectively} massive particle. In this context, it has
long been observed that superradiance could be due to photons propagating in a plasma~\cite{Press:1972zz}, wherein they are effectively described by a Proca equation $\nabla_\alpha F^{\alpha \beta}=\omega_{\rm pl}^2 A^\beta$. This idea has been used in~\cite{Pani:2013hpa} for constraining the contribution  to dark matter of spinning primordial BHs.

This paper is founded on a simple numerical observation which, nonetheless, has not to our knowledge been made before: the numerical matching of equation (1) is satisfied for stellar mass BHs and the photon plasma mass applicable within the diffuse environment of galaxies or galaxy clusters.
More precisely, the effective photon mass in a plasma is given by
\bea
\omega_{pl} = \left(4 \pi \alpha \frac{n_e}{m_e}\right)^{1/2} =1.2\cdot10^{-12} \sqrt{\frac{n_e}{10^{-3}\  {\rm cm}^{-3}}}~{\rm eV} \nonumber \, ,
\eea
where $\alpha = \frac{e^2}{4 \pi c \epsilon_0 \hbar}$ is the fine structure constant.
The diffuse galactic free electron density is described by (for example) the NE2001 model \cite{Cordes:2002wz}.
It varies from around $n_e \sim 10 \ {\rm cm}^{-3}$ in the inner $\sim 50 \ {\rm pc}$ of the galaxy,
through around $4 \ti 10^{-2} \ {\rm cm}^{-3}$ near our solar system, falling away rapidly as one moves vertically away from the disk ($e.g.$ see \cite{Gomez:2001te, Cordes:2002wz,Gaensler:2008ec}).
In the intracluster medium within galaxy clusters, $n_e \sim 3 \ti 10^{-2} \ {\rm cm}^{-2}$ near the centre of cool core clusters,
with $n_e \sim 3 \ti 10^{-3}\ {\rm cm}^{-3}$ at the center of a bright non-cool-core cluster such as Coma, and with $n_e \sim 10^{-3}\ {\rm cm}^{-3}$
as a `typical' value within a galaxy cluster. In deep intergalactic space,
$n_e$ reaches $\sim 10^{-7} \ {\rm cm}^{-3}$.

It follows that within a typical `empty' interstellar environment within a galaxy or galaxy cluster, the photon has an effective mass
given by $\omega_{\rm pl}$ which is in the range $10^{-12} - 10^{-10} \ {\rm eV}$.
Electromagnetic modes with $\omega < \omega_{pl}$ are therefore unable to propagate, suggesting that, when confined by the plasma near to the BH, they will undergo superradiant amplification.
Thus, this observation seems to imply that \textit{bare}, highly spinning stellar mass Kerr BHs
in galactic or intracluster environments would be unstable to superradiant amplification of low-energy electromagnetic modes that are confined by the ambient galactic plasma.

The general nature of the superradiant instability leads to a formation of a boson cloud around Kerr BHs.
In the scalar (vector) case this has been shown to comprise up to $\mc{O}(20 \%)$ [$\mc{O}(10 \%)$] of the mass
of the original BH~\cite{Brito:2014wla} (\cite{East:2017ovw}). A photon-induced instability
may then lead to a rapid transfer, on millisecond timescales, of energy from the mass of the BH into a photon cloud bound by
the gravitational potential of the BH with potentially observable consequences.

\section{The Superradiant Instability}
The linearity of the massive Klein-Gordon or Proca equations (mass $\mu$), as test fields on the Kerr background, allows us to study a single mode, in the frequency domain, to understand the basic features of the superradiant instability's \textit{linear phase}, triggered by these fields. Then,  each Fourier mode of the scalar/Proca field is characterized by a set of quantum numbers, including the angular ones, $\ell,m$, and a complex frequency, $\omega=\omega_R+i\omega_I$. Analytic computations of $\omega_R,\omega_I$ can be done in the asymptotic regimes $M\mu\ll 1$ or $M\mu\gg 1$. For the real part, to lowest order in $M\mu$, one finds a hydrogen-like spectrum,  since, asymptotically, the field experiences a $1/r$ potential on the Kerr geometry~\cite{Furuhashi:2004jk,Rosa:2011my,Pani:2012bp}.  Here we are mostly interested in the imaginary part, whose inverse determines the instability timescale $\tau_{SR}=w_I^{-1}$.  For  $M\mu\ll 1$, the wave equation can be
dealt with using a matching method and one finds that $M\omega_I$ is suppressed as $(M\mu)^a$, where $a=9$ for scalars~\cite{Detweiler:1980uk} and $a=4\ell+5+2S$ for vectors~\cite{Pani:2012bp}, with $S=0,\pm1$ being the mode's polarization. For $M\mu\gg 1$, which requires large quantum numbers $m,\ell \gg 1$, the wave equation is amenable to the WKB approximation and one finds that $M\omega_I$ is suppressed as $e^{-b M\mu}$, where $b=1.84$ for scalars~\cite{Zouros:1979iw}. In the optimal regime for the instability $M\mu\ \sim1$, the information must be obtained numerically.  Then, for both the scalar case~\cite{Dolan:2012yt} and vector case \cite{Witek:2012tr}, the largest growth rates are obtained for the $\ell=m=1$ mode on an almost extremal Kerr BH $a\sim0.99$ and for $M\mu\sim 0.4$ and they are:
$\tau_{SR}^{scalar}\sim 10^7 M \  , \tau_{SR}^{Proca}\sim 10^3 M.$
These results show that the timescale for the vector instability is far more rapid than that for the scalar instability. The strongest instability for the
vector case can be expressed as
\be
\tau_{SR}^{Proca} \sim 10^{-2} \gamma_{-11}^{-1} \left( \frac{M}{M_{\odot}} \right) {\rm s} \sim 5 \ti 10^{-4} \left( \frac{M}{M_{\odot}} \right) {\rm s} \ ,
\ee
using $\gamma_{-11} \sim 20$~\cite{Pani:2012bp} (see also~\cite{Endlich:2016jgc}). This number was estimated in the small rotation regime but the numerical results of~\cite{Witek:2012tr} show that for highly spinning Kerr BHs the corresponding instability timescales differ only by a factor of 2 (see also~\cite{East:2017mrj}).
Thus for $M\sim \mc{O}(1 - 100) \ M_{\odot}$, the corresponding timescales are $\tau_{SR}^{Proca} \sim 0.5-50$ ms.

Beyond the linear regime, the evolution for either the scalar or vector case is not precisely known. Partial results using different approximations suggest 1) the superradiant growth may extract a considerable fraction of the original BH's mass (up to $\sim$20\% in the scalar case~\cite{Brito:2014wla} and up to $\sim$10\% in the vector case~\cite{East:2017ovw})). For a real field, the cloud will
necessarily eventually decay by emission of gravitational and (for a photon cloud) electromagnetic waves; 2) before (or instead of a smooth) saturation, non-linear phenomena may trigger an explosion akin to a bosenova,
terminating the superradiant growth. This phenomenon has been studied for a self-interacting scalar field via numerical simulations~\cite{Yoshino:2012kn,Yoshino:2013ofa,Yoshino:2015nsa}. Depending on the initial data, different outcomes are possible, including the ejection of part of the bosonic cloud away from the BH, associated to the generation of higher frequency modes. Another part of the cloud falls towards the BH, terminating the energy extraction. The instability may, however, kick in again, creating recurrent cycles~ \cite{Arvanitaki:2010sy,Yoshino:2012kn}.

In the vector case, photon self-interactions can arise either from
the induced four-photon interaction in the low energy limit of QED (the Euler-Heisenberg term~\cite{Heisenberg:1935qt}) or from the effect of the electric and magnetic fields of the photon cloud
on the surrounding plasma.
In the presence of photon self-interactions, which become important once the photon cloud grows beyond some threshold, a similar phenomenon should arise for the Proca case, causing dissipation of some part of the cloud, and re-absorption of the remaining part into the BH.
Once this happens, the instability can become operative again.

As galactic plasma frequencies correspond to long-wave radio frequencies ($10^{-10} \  {\rm eV} \equiv 24 \ {\rm kHz}$),
such a ``bosenova" would be observed as an electromagnetic burst in radio frequencies, which we dub a ``\textit{radionova}".

\section{Astrophysical scenario: formation of a bare BH}
%
Our above usage of the diffuse galactic plasma mass would only apply for photons around a BH if the latter is ``bare". In the presence of a substantial accretion disk,
or for the case of a stellar-BH binary in which the star provides a continual source of matter, the local free electron density is likely to be significantly
higher than for the diffuse galactic plasma. For this reason, one would not expect a
 BH spun up by an accretion disk to be unstable whilst the accretion disk remains; the local plasma mass is
large and for $M \omega_{\rm pl} \gg 1$ the superradiant instability timescale is exponentially suppressed with $e^{+M \omega_{\rm pl}}$.
How can we ensure the presence of a bare BH?

A conservative way is to consider a BH produced following a binary BH merger. We know that such mergers
 occur in the universe and are capable of producing BHs with (dimensionless) spin $\sim 0.7$ \cite{Abbott:2016blz,Abbott:2016nmj}.
Mergers of BH binaries can lead to substantial kicks being imparted to the resulting BH, with recoil velocities of up to $5000 \  {\rm km \ s}^{-1}$~\cite{Gonzalez:2006md,Campanelli:2007cga,Lousto:2011kp,Gerosa:2014gja}.
Such a hypervelocity BH will be thrown clear of its formation nest and will in time leave its parent galaxy. Travelling at a hypervelocity speed,
it will not grow an accretion disc from its local environment and will eventually pass into intergalactic space.
If such a hypervelocity $\mc{O}(1 - 100) \ M_{\odot}$ BH originates in a dense highly ionised
environment where $\omega_{\rm pl} \gtrsim 10^{-9}\  {\rm eV}$ (corresponding to $n_e \gtrsim 10^3 \, {\rm cm}^{-3}$),
then on its passage to intergalactic space (where $\omega_{\rm pl} \sim 10^{-13} - 10^{-14} \ {\rm eV}$) it experiences
a constantly decreasing local electron density.
It will then pass through a region
where superradiance is maximally effective, triggering the instability.

A hypervelocity kick guarantees that a product BH escapes its local environment. However, this may not be necessary.
The superradiant instability is most efficient at high BH spin. If the product of
a binary BH merger is more highly spinning than either of the progenitors, then even without any change
in $\omega_{\rm pl}$ the product may be capable of triggering the superradiant instability while the progenitors were not.
For instance, the LIGO collaboration has observed in GW150914 the production of a $\sim 60 M_{\odot}$ BH with spin $\sim 0.7$ \cite{Abbott:2016blz}.
\ For such a BH the superradiance instability would
be maximally efficient for $\mu \sim  10^{-12} \, {\rm eV}$. This is only slightly smaller than a typical
galactic plasma frequency of $\omega \sim 10^{-11} \, {\rm eV}$, and so this BH may have undergone the superradiant instability.

Finally, it may even be the case that the instability could be operative for a highly spinning BH formed directly from a supernova; this would require
the supernova to blow off the original stellar material to a sufficient extent that the local electron density was reduced to close to a galactic value
of $\sim 10^{-2} \ {\rm cm}^{-3}$.

\section{Astrophysical Phenomenology}
If astrophysical Kerr BHs are unstable to a plasma-induced superradiant instability, leading to the formation of photon clouds,
what is the
resulting observational phenomenology? A fully rigorous discussion would require
analytic and numerical simulations of the instabilities, which are beyond the scope of this paper. We provide here a qualitative enumeration of possibilities, while
emphasising that the precise endpoint of the instability is not known.

The starting point of the instability is exponential growth of a photon cloud around the BH.
The energy of these individual photons is $\sim 10^{-10} \ {\rm eV}$.
For solar mass BHs the instability timescale for a massive vector is $\sim 1 - 10 \  {\rm ms}$. As
the cloud grows exponentially, maximal
energy extraction will occur in the last doubling period.
The instability then leads to the extraction of a large amount of energy (in the most optimistic case up to
$\sim 10^{46}  \  {\rm J}$) from the BH and into the photon cloud over an $\mc{O} (10 \ {\rm ms})$
timescale. 

Despite the high energies involved, this corresponds to a deeply infrared phenomenon -- the photon cloud is confined by the surrounding plasma, and
the large energies occur through an $N \gg 1$ occupation number of the bound state of the
massive photon in the gravitational potential.

As there are no stationary states of a BH with a bosonic cloud for real fields, the energy in the photon cloud must be dissipated. For scalar clouds, this can occur
via a bosenova, where some fraction of the cloud escapes to infinity as high-frequency modes while the remainder is re-absorbed by the BH.
What happens with a vector field?

There are two forms of vector field interactions that will become important at high field strength. First, there is the non-linear Euler-Heisenberg interaction
\be
\mc{L}_{E-H} = \frac{2 \alpha^2}{45 m_e^4} \left[ \left({\bf E}^2 - {\bf B}^2 \right)^2 + 7 \left( {\bf E} \cdot {\bf B} \right)^2 \right].
\ee
This is a four-photon interaction that is induced in QED by electron-positron loops, and
becomes important for background field strengths comparable to the mass-energy of the electron ($B \sim 10^{13}\  {\rm G}$), and which
 for $B \gtrsim 10^{16}\  {\rm G}$ becomes dominant.

Secondly, there is the reaction of the field on the surrounding plasma.
Associated to the growing photon cloud are growing electric and magnetic fields. The photon cloud only grows because it involves a mode confined by the surrounding plasma.
Large ${\bf E}$ and ${\bf B}$ fields will interact with the plasma, accelerating the constituent particles and
altering its local properties. This may modify the shape and growth of the photon cloud,
possibly acting as an endpoint for the instability.

These self-interactions and backreaction effects may cause the growth of the cloud to terminate while the energy of the photon cloud is still small compared
to the original BH. In the case of non-linearities from the Euler-Heisenberg term, for a solar mass BH
a magnetic field of $\sim 10^{16} \ {\rm G}$ extended across one Schwarzschild radius corresponds to a photon cloud
containing $\sim 10^{-6} $ of the BH mass (this fraction grows quadratically with the BH mass).
Both forms of interaction also have the potential to lead to a radionova, where some fraction of the photon cloud collapses back into the BH and
some fraction is expelled to infinity as higher frequency modes -- although we again stress
that a quantitative analysis requires full simulations of the instability.

If plasma-induced superradiance occurs for astrophysical BHs, it should be a phenomenologically important phenomenon.
In this respect there is a speculative connection to Fast
Radio Bursts (FRBs)~\cite{Lorimer:2007qn} (a recent review is~\cite{Katz:2016dti}). These are mysterious,
highly energetic, transient signals observed at radio frequencies $f \sim 1 \ {\rm GHz}$, with durations of several milliseconds and at cosmological
distances. The emitted energy at radio frequencies is $\sim 10^{32} - 10^{33} \  {\rm J}$ across the burst, a rate of $\sim 10^{41} - 10^{42} \ {\rm erg\ s}^{-1}$.
One burst has been observed to repeat \cite{Spitler} and has recently been located to a dwarf galaxy at redshift $z$=0.19 \cite{Chatterjee, Marcote, Tendulkar}.
No definitive high-energy counterparts have been observed, although in one case there is a claimed gamma ray counterpart at 3.2 sigma significance \cite{161103139}.

Such plasma-induced superradiance would have three of the necessary features to describe FRBs.
First, it involves large energy releases, due to the exponential growth in the energy stored in the photon cloud.
Second, when the superradiance matching condition Eq. \ref{superradiancecondition} holds, the dynamical timescales for massive vectors
is of order milliseconds, fitting the temporal extent of FRBs.
Finally, superradiance is not catastrophic, and so is compatible with repeating Fast Radio Bursts. As one FRB has been
observed to repeat \cite{Spitler}, at least some FRBs must arise from non-destructive processes that do not destroy the original source.

A scalar bosenova has this repeating property -- the reabsorption of the bosonic cloud into the BH returns it to a higher spinning state, and allows
for multiple occurrences of the superradiant instability \cite{Arvanitaki:2010sy,Yoshino:2012kn}. Despite this suggestive evidence from the scalar case, the occurrence of repeated radionovas is a question that can only be
answered by detailed numerical simulation with a self-interacting Proca field on the Kerr background, which seem possible with current technology~\cite{Zilhao:2015tya,East:2017ovw}.

We note that one potential problem with this is that
 the GHz radio frequencies in FRB detections ($E \sim 5 \ti 10^{-6} \ {\rm eV}$) are a few orders of
magnitude more energetic than the characteristic mass scale for the superradiant
instability of $E \sim 10^{-10}\  {\rm eV}$ for a solar mass BH (one possible
way to imagine increasing photon energies is by acceleration of electrons in the large ${\bf E}$ and
${\bf B}$ fields of the photon cloud, following by inverse Compton upscattering of photons).

The above focussed on transient signals arising from electromagnetic interactions as the superradiant instability reaches an endpoint.
It may also be the case that there is a long-lived solution of a bound photon cloud around a BH, in which the photon cloud only loses
energy slowly via either gravitational or electromagnetic wave emission -- see \cite{Baryakhtar:2017ngi} for a study of this type of scenario for dark photons.

If this occurs, there are other possible origins of astrophysical signals.
Suppose the original bare Kerr BH is realised as a highly kicked product of a BH binary merger, and is travelling rapidly through a galaxy.
In doing so it may pass through a shock front into a region where the free electron density jumps (either upwards or downwards).
A discontinuity in electron density could arise at the boundary of an expanding supernova bubble or an
HII region. From the perspective of the superradiant instability, this corresponds to a sudden increase (or decrease) in the effective
photon mass.

If the BH moves into a region of lower $n_e$ (and so reduced $\omega_{\rm pl}$) then the plasma no longer provides a barrier to the
propagation of the bound photon states to infinity. In this case the energy stored in the bound photon states can then suddenly escape as
low energy radio emission at kHz frequencies.

Alternatively, moving into a region of higher $n_e$ would cause a sudden increase in $\omega_{\rm pl}$. In this case the Compton wavelength of the photon
would suddenly decrease, causing much of the photon cloud to be absorbed by the BH. As the increased photon mass would make the
photon cloud no longer able to support itself against the gravitational attraction of the BH, the cloud would undergo gravitational collapse. As with collapsing stars,
one would expect this collapse to also be explosive, resulting in a burst of radiation as some of the photons are expelled to infinity.

Both scenarios appear realisable. If a highly kicked BH binary merger occurs in the center of a galaxy, in a region of high $n_e$, then its
subsequent passage to intergalactic space would take it through regions of ever lower $n_e$, allowing first the triggering of the superradiant instability
and later the escape of the photon cloud as $n_e$ decreased further. Alternatively, there will also be mergers where the product BH is kicked towards
the center of a galaxy, and so passes through progressively larger values of $n_e$ that will cause collapse of the photon cloud.

\section{Limitations of the Proca model?}
%
A possible concern for the scenario presented herein is how well the Proca model can describe the interactions of photons with the plasma in the \textit{vicinity} of the BH.
The attractive nature of the BH will cause the local plasma density to be larger than the diffuse value, and so there will be a gradient
in the plasma mass rather a constant value. However, superradiance occurs when modes are confined to the vicinity of the Kerr BH
and it is the case here that the asymptotic plasma mass is unaffected, and so low-frequency modes will always be confined.
For the BH in Sgr A*, at the gravitational centre of the Milky Way, the plasma frequency at a few Schwarzschild radii is
only around three orders of magnitude larger than the diffuse value in the inner galaxy \cite{11120026}, and for a newly formed BH with no accretion history the difference
should be much smaller.

\section{Conclusions}
%
The main point of this paper is simple: the effective plasma-induced mass of the photon in either galactic or intracluster plasmas
is in the correct range to trigger the superradiant instability of bare highly spinning solar mass Kerr BHs.
While the precise non-linear evolution and endpoint is not known,
if this instability occurs it is likely to have interesting observable consequences, causing
highly spinning astrophysical BHs to transfer large quantities of energy to the electromagnetic field over
millisecond timescales. As the superradiant instability gives rapid energy release in the radio band, it also represents a speculative
origin of the mysterious Fast Radio Bursts.

\bigskip

{\bf Acknowledgements.} We thank the organizers -- especially Hideo Kodama --
and participants of the 5th UT Quest Workshop at the Yukawa Institute for Theoretical Physics for a stimulating environment, where
this paper was conceived. We also thank V. Cardoso, S. Dolan, E. Berti, P. Pani and J. G. Rosa for comments on this manuscript. This project is funded in part by the European Research Council starting grant `Supersymmetry Breaking in String Theory' (307605).
C.H. acknowledges funding from the FCT-IF programme.  This project has received funding
from  the  European  Union's  Horizon  2020  research  and  innovation  programme  under  the H2020-MSCA-RISE-2015 Grant No.   StronGrHEP-690904, the H2020-MSCA-RISE-2017 Grant No. FunFiCO-777740  and  by  the  CIDMA  project UID/MAT/04106/2013.
The authors  would also  like  to  acknowledge networking support by the COST Action GWverse CA16104.
We thank Sam Dolan, Paolo Pani and Joao Rosa for comments on the draft.

\end{document}